\documentclass{article}
\usepackage{arxiv}

\usepackage[utf8]{inputenc} 
\usepackage[T1]{fontenc}    
\usepackage{hyperref}       
\usepackage{url}            
\usepackage{booktabs}       
\usepackage{amsfonts}       
\usepackage{nicefrac}       
\usepackage{microtype}      
\usepackage{lipsum}
\usepackage{amsmath}
\usepackage{graphicx}
\usepackage{tabu}

\usepackage{subfig}

 \title{\textbf{Faked Speech Detection with Zero Prior Knowledge}}

\author{
  Sahar~Al~Ajmi, \hspace{2mm}\href{https://orcid.org/0000-0001-5216-6019}{\includegraphics[scale=0.06]{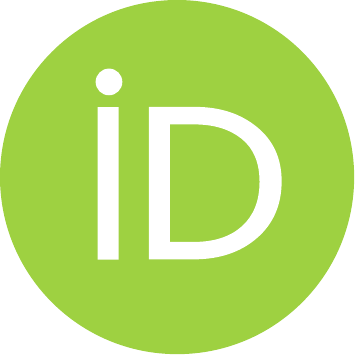}}Khizar~Hayat\thanks{Corresponding author}, \hspace{2mm}Alaa~M.~Al~Obaidi, \hspace{1mm}Naresh~Kumar, \hspace{1mm} Munaf~Najmuldeen \\
  University of Nizwa,\\
  Sultanate of Oman \\
   \texttt{Saharalajmi4@gmail.com,\{khizar.hayat,alaamohammed,naresh,munaf\}@unizwa.edu.om}\\
   \AND
   \href{https://orcid.org/0000-0003-3458-0552}{\includegraphics[scale=0.06]{orcid.pdf}} Baptiste~Magnier \\
    Euromov Digital Health in Motion, \\
    Univ Montpellier, \\
    IMT Mines Ales, Ales, France \\
    \texttt{baptiste.magnier@mines-ales.fr}\\
}

\renewcommand{\shorttitle}{Digital Voice Mimicry Forensics}

\begin{document}
\maketitle
\begin{abstract}\normalsize
Audio is one of the most used ways of human communication, but at the same time it can be easily misused to trick people. 
With the revolution of AI, the related technologies are now accessible to almost everyone, thus making it simple for the criminals to commit crimes and forgeries. 
In this work, we introduce a neural network method to develop a classifier that will blindly classify an input audio as real or mimicked; the word 'blindly' refers to the ability to detect mimicked audio without references or real sources.  
We propose a deep neural network following a sequential model that comprises three hidden layers, with alternating dense and drop out layers. The proposed model was trained on a set of $26$ important features extracted from a large dataset of audios to get a classifier that was tested on the same set of features from different audios. 
The data was extracted from two raw datasets, especially composed for this work; an all English dataset and a mixed dataset (Arabic plus English)\footnote{The dataset can be provided, in raw form, by writing an email to the first author}. 
For the purpose of comparison, the audios were also classified through human inspection with the subjects being the native speakers. 
The ensued results were interesting and exhibited formidable accuracy, as we were able to get at least $94\%$ correct classification of the test cases, as against the $85\%$ accuracy in the case of human observers.
\end{abstract}


\section{Introduction}
\label{sec:Intro 1}
\begin{quote}
\textit{By the time the truth arrives, lies have already destroyed the countryside.}
    \begin{flushright}
        -- A Pushtu proverb
    \end{flushright}
\end{quote}

\begin{quote}
\textit{The world today is such that whatever people don’t know about you, they create. They piece together what they hear from rumours, social media as well as assumptions \& lies. It spreads really fast by the touch of a button...}
    \begin{flushright} 
        -- Mufti Menk  
    \end{flushright}
\end{quote}  
This is an age of mendacity, backed up by digital art; lies are spreading lot faster than truth, especially through social media platforms. 
Fake multimedia content are mainly present in the form of texts, audios, images and videos, but now we have what is referred to as ``deep fakes'', thanks largely to the mushroom growth of deep learning algorithms. 
High quality fake audios/videos are commonplace and, sometimes,  threatening to ruin lives. 
Add to it statistics, and you would find keen recipients everywhere in the world, in this post-truth era, from almost every age group and with every educational background. 
In the words of Mark Twain, "There are three kinds of lies: lies, damned lies, and statistics."
At the same time, we cannot undermine the importance of circumstantially correct and available media, nor can we decrease the importance of statistics. 

Audiovisual media is one of the most important instruments to highlight not only atrocities/genocides across the globe but also fixing the responsibility of a given crime on the guilty. 
Its importance in a court of law, as well as the court of the people, cannot be underestimated.
Hence, we cannot trash altogether important means of evidence merely on the suspicion that these may be fakes. 
There has to be criteria to evaluate the veracity of a given multimedia content; be it an audio, an image or a video.

Digital Multimedia forensics has obtained a lot of attention during the past decade. 
Most of the work has been generally focused on Image forgery detection \cite{1} with copy/move forgery detection taking a lion share \cite{2}. 
In comparison, digital audio forgery detection has not got that much attention; it is important in multimedia forensics to ensure the authenticity and integrity of the data in hand before handling it as an evidence. 
The primary focus of audio forensics is to establish the integrity of the audio, whether it is real or fake, and also identify the real person/s who is/are talking. 
The goals may be diverse, ranging from using it as evidence in a court of law to preempting social media or paparazzi rumors. 
Digital impersonation encompasses ways to produce a speech that may deceive people/machines into identifying it with a legitimate and authentic source, which could result in a social or economic loss. 
Indeed, one important aspect of audio impersonation deals with mimicking someone’s voice to attribute to him/her something which was never said in one go in a context that may be uncomfortable.

Due to the availability of skilled voice actors, detection of non-machine human audio mimicry is still a challenge, especially if it is \textit{blind}, i.e., the partial or complete absence of the authentic recorded voice of the person to whom the audio is attributed.   
In other words, the ability to detect mimicked audio without the existence of any target reference; original, mimicked, or impersonated. 
This difficulty, to blindly distinguish between real audios from mimicked audios, forms the basis of the research problem being addressed in this work. 
To be specific, our research problem is characterized by the following question:\\
\begin{flushright}
\centering
        \textit{Can we come up with a method to blindly classify a given audio as human mimicked (faked) or otherwise (real) in the absence of any other genuine or disguised audio by the speaker?}
    \end{flushright}
A supplementary question could be related to \textit{the dependence on the spoken language}. Our underlying assumption is that \textit{the system has never encountered the purported speaker before}. 

The recent advances in Machine Learning (ML) - especially neural networks and deep learning - can be exploited in forgery detection in audiovisual data. 
With the potential amount of available data being huge, deep learning can be the best way to classify it. 
The idea is to extract important features from a lot of audios and feed it to a neural/deep network that will help the model to learn how to identify the real audios from faked audios. 

The rest of the paper is arranged as follows. 
Section~\ref{sec:speech} briefly describes the background speech processing concepts needed for the comprehension of this article. 
The related work from literature is outlined in Section~\ref{sec:rel} which is followed by Section~\ref{sec:dset} to introduce our dataset for this work. 
Section~\ref{sec:method} explains our methodology which is then trained and tested on the dataset presented in Section~\ref{sec:results} that analyses all the ensued results. Section~\ref{sec:ref} outlines the benchmarking results.  
Finally, Section~\ref{sec:concl} concludes the paper. 

\section{Speech Processing}\label{sec:speech}
It is important to know the difference between audio and speech. 
While an audio is a waveform data in which the amplitude changes with respect to time, speech is the oral communication \cite{b} and pertains to the act of speaking and expressing thoughts and emotions by sounds and gestures. 
The human brain processes and analyses everything around to help the body with the right response or reaction, and it does the same for the voices \cite{a}. 
To be able to hear and process a voice, the inside of a human ear is equipped with small hair of various sizes; some are short and respond to resonate with low frequency voices while others are long that resonate with the high frequency voices. 
Each of these hairs is connected to a nerve that carries a signal to the brain for processing \cite{c}. 
    
An audio signal is a representation of sound in function to the vibration of sound that is audible to the human ear \cite{d}. 
Audio frequency \cite{e} is the periodic variation of sound, with the human audible frequency being 20 Hz to 20 kHz \cite{f}. 
For a machine, the processing of an audio is different from humans. 
In order for the machine to get a sound it should have the needed devices that are able to record and save the audio in machine processable formats like the well-known mp3, WMA, WAV, or others.
     
Features from speech signals can be broadly classified as temporal and spectral features. The temporal features are time domain features having simple physical interpretation and easy to compute. Examples are signal's energy, maximum amplitude, zero crossing rate, minimum energy etc~\cite{spc}. 
The spectral features, on the other hand, are frequency-based features that are extracted after passing the time domain signal to the frequency domain using Fourier or other similar transforms. 
Examples are  frequency components, fundamental frequency, centroid, spectral flux, spectral density, roll-off etc~\cite{spc}. 
In the context of audio signals such features may be helpful in the identification of pitch, notes, rhythm and melody etc. 
 
\emph{Spectrum} and \emph{cepstrum} are two important frequency-based concepts in audio processing. A spectrum is mathematically a Fourier transform of a signal which converts a time-domain signal into frequency domain~\cite{Singh2019}, i.e. spectrum is the audio signal in frequency domain. 
A cepstrum is the log of the magnitude of the spectrum followed by an inverse Fourier transform. That's why its domain is neither frequency nor the time; its domain is called \emph{quefrency}~\cite{Singh2019}. Cepstrum can be said of as a sequence of numbers that characterize a frame of speech~\cite{g}. 
Since the Fourier transform is a linear operation, so is consequently the cepstrum; the spectrums of the wavelet and reflectivity series are additively combined in the cepstrum~\cite{Hall2012}. 
Following are some important features~\cite{kotha2020} exploited in speech processing:
%
\begin{itemize}
\item \textit{Zero crossing rate:}  It indicates the number of times the value of the signal changes between positive and negative and vise versa. 
It is also used to measure the noise in a signal, and it usually gives high value in case of a noisy signal \cite{zero}.
\item \textit{Spectral centroid:} It is a feature based on frequency which indicates the location of the center of mass of the spectrum. 
In audios, it is known as a good predictor of `` brightness'' of a sound \cite{sc}.
\item \textit{Spectral roll off:} This feature is used to differentiate between the harmonic sound (below roll off) and the noise sound (above roll off). 
It is known as the energy spectrum under a specific percentage that is defined by the used (85\% by default) \cite{sro}.
\item \textit{Spectral bandwidth:} The difference between the higher and lower frequencies in a group of continuous frequencies.
\item \textit{Chroma:}  
This representation for audio where the spectrum is divided onto 12 bins representing the 12 distinct semitones (or chroma) of the musical octave.
\item \textit{Root Mean Square Energy (RMSE): } RMSE represents the energy of the signal, and shows how loud the signal is \cite{energy}.
\item \textit{Spectral flux:} It measures how quick the power spectrum of a signal is changing, and it is calculated by comparing the changes of the power spectrum between one frame and the frame before it\cite{specflx}.
\item \textit{Spectral density:} It is the measure of signal's power content against frequency \cite{spcden}.
%
\item \textit{Cepstral Features:}
These are, as stated above, \emph{quefrency} domain features with the following being considered important:
\begin{itemize}
\item \underline{Mel Frequency Cepstral Coefficients (MFCCs~\cite{Davis1980,Mermelstein1976}):} MFCCs are widely used features for speech recognition.  
The Mel-frequency	scale	represents	subjective or perceived pitch as its	construction is based on pairwise comparisons of	sinusoidal	tones. 
The conversion between Hertz ($f$) and Mel frequencies ($m$) can be generalized as:
\begin{equation}
	m=2595 \cdot \log \left(1+\frac{f}{700}\right),
\end{equation}
\begin{equation}
	f=700 \cdot (10^{m/2595} - 1).
\end{equation}
MFCCs are obtained by applying a short time Fourier transform to window-based slices from the audio signal, followed by calculating the power spectrum and consequently filter banks (triangular in shape). 
The filter bank coefficients are highly correlated and one way to de-correlate them is by applying a Discrete Cosine Transform {DCT} to get a compressed representation in the form of {MFCC}. 
Typically, MFCC 2-13 (i.e. 12 coefficients) are kept, and the rest are discarded \cite{ppp}.
\item \underline{Gammatone Frequency Cepstral Coefficients (GFCCs)~\cite{Zhao2013}:} used in a number of speech processing applications, such as speaker identification. 
A Gammatone filter bank approximates the	impulse	response	of	the	auditory	nerve	fiber	thus emulating human hearing and its shape can be likened to a \emph{Gamma function} ($e^{-2\pi (f_c) b t}$) modulating the \emph{tone} ($\cos (2\pi f_{c}t+\phi)$) \cite{12ss}:
\begin{equation}
	g(t) = at^{n-1}e^{-2\pi (f_c) b t}\cos (2\pi f_{c}t+\phi)
\end{equation}
Where $a$ is	peak	value,	$n$ the order of the filter,	$b$ the	bandwidth,	$f_c$	the	characteristic frequency and $\phi$	is	initial	phase.	$f_c$ and $b$ can be derived from Equivalent Rectangular Bandwidth (ERB) scale, using the following equation~\cite{Jeevan2017}:
\begin{equation}
	\text{ERB}(f_c) = 24.7 \cdot \left(4.37 \cdot \frac{f_c}{1000} + 1\right)
\end{equation}
\begin{equation}
	b = 1.019 \times \text{ERB}(f_c)
\end{equation}
For GFCC, FFT treated speech signal is multiplied by the Gammatone ﬁlter bank, reverted back by IFFT, noise is suppressed by decimating it to $100\ Hz$ and rectiﬁed using a non-linear process. The rectification is carried out by applying a cubic root operation to the absolute valued input
~\cite{Jeevan2017}. Approximately, first 22 features are called GFCC and these may be very useful in speaker identification. For a concise comparison on MFCC and GFCC, the reader can further consult~\cite{Zhao2013}. 
    \end{itemize}
Linear Prediction Cepstral Coefficients (LPCCs) and Linear Prediction Coefficients (LPCs) were the main features used in automatic speech recognition before MFCC specially with Hidden Markov Model (HMM) classifiers \cite{g}. 
Some other important Cepstral features are Bark Frequency Cepstral Coefficients (BFCCs) and
Power-Normalized Cepstral Coefficients (PNCCs).
\end{itemize}

\section{Related work}\label{sec:rel}
\subsection{The Spectrum of Voice Impersonation}
The output of \textit{voice impersonation} must be convincing, both to humans and machines, in being naturally uttered by the target speaker.
This requires mimicking the signal qualities, like pitch, as well as the speaking style of the target~\cite{Gao2018}. 
In this age of deep fakes, seamless machine-based impersonation is a reality. 
The method in~\cite{Gao2018} relies on using a neural network-based framework that uses Grifﬁn-Lim method~\cite{Griffin1984} which can learn to mimic a person’s voice and style and then produce a voice that mimics the persons’ voice. 

\subsubsection{Voice cloning} Cloning technologies can learn the characteristics of the target speaker and utilize prepared models to mimic a person’s voice from only a few sound samples. 
The developments in cloned speech generation technologies can create a fake machine speech that is similar to the real voice of the target speaker~\cite{Malik2019}. 
There have been research efforts focusing on how to detect this kind of audio and how to enable systems to recognize them, such as the anti-Spoofing works~\cite{Gomez-Alanis19a,Yamagishi2021,Gomez-Alanis19,Tak2021}. In this context, the ASVspoof series of challenges are an important mention in dealing with spoofing in  automatic speaker verification systems. ASVspoof baseline systems are based in modern deep learning architectures and having deep- and handcrafted features. although related, the problem we are dealing here is yet to be taken up in these challenges, to the best of our knowledge at least. 

\subsubsection{Voice disguise} Voice disguise refers to altering one's voice deliberately to conceal one's identity. 
Impersonation is concerned with a voice disguise aimed at sounding like another person who exists~\cite{Delvaux2017}. 
While impersonation may be easily detectable, it is a hard task to trace back a disguised voice, of presumably a person who never existed, to the original speaker.
A study in~\cite{Wagner1999} reaffirms the importance of phonetically trained specialists in subjective voice disguise identification after an untrained audience failed to identify known speakers in case of falsetto disguise. 
Another study along similar lines~\cite{Delvaux2017}, reports that naive listeners can better distinguish between an impersonator and a target rather than identifying voice disguise. 
Readers are recommended a review of similar studies~\cite{Perrot2007}. 
The system proposed in~\cite{Chen2017} relies on the magnetic ﬁeld produced by loudspeakers to  detect machine-based voice impersonation attacks. 
The reported results in combination with a contemporary system against human impersonation attacks, are incredible, viz. $100\%$ accuracy and $0\%$ Equal Error Rate (EER)~\cite{FURUI2010}.

\subsubsection{Human Mimicry}
Human attempted voice impersonation (or voice imitation) is mimicry of another speaker’s voice characteristics and speech behavior~\cite{hautamaki2013} without relying on computer related spoofing; in fact, ruling out the quest for technical artifacts in the suspected audio.
The focus is mainly on voice timbre and prosody of the target~\cite{hao19}. 
Being a ``technically valid speech'', mimic attacks may not be detectable, especially in Automatic Speaker verification (ASV) environments. 
A professional impersonator is likely to target all lexical, prosodic~\cite{farrus2008} and idiosyncratic aspects of the subject speaker; exaggeration may be inevitable~\cite{hautamaki2013}.  
The study reported in~\cite{patil2009} states that speech patterns, pitch contours, formant contours, and spectrograms etc. from speech signals of maternal twins are at least almost identical, if not exactly the same. 
Hence, even a mere verification may be a difficult task in the case of identical twins. 
Therefore, more exploration of discriminating speech features is needed, as suggested about half a century ago~\cite {rosenberg1976}. Even a paternal twin may be hand, as in a recent incident related to phone banking~\cite{HSBC17,BBC2017}, a non-identical twin mimic the voice of his brother, a BBC reporter, to deceive the system~\cite{Finextra2017}. 
The literature contains many such incident of fraud~\cite{jain2002}. 

\subsection{Detecting Human Mimicry}
Even extensive works, like~\cite{Zakariah2018,masood2023}, do little to touch the subject of impersonation, especially the human mimicry, i.e., mimicking someone's voice to attribute to him/her something which was neither ever said nor uttered in one go. 
Here, we are talking of an audio that has never been tampered; other than the usual pre-processing, ﬁltering and compression etc.  
As of datasets, there exist quite a few forensic databases~\cite{Kraetzer2007}, but even these don't touch the aspect of impersonation in the form of human mimicking. 

The problem of human mimicry may be the earliest one addressed in the literature and can be traced back to as far as 1970s. 
For example, an old study~\cite{Reich1976} employed four professional experts to identify voice disguises from the spectrogram of two sentences uttered by a sample of 30 subjects (15 reference + 15 matching) in undisguised as well as five disguised modes. 
Even without any disguise, the experts could go as far as $56\%$ accuracy in matching the speakers.
To classify speakers, another early days' simulation in~\cite{Wolf1972} uses such parameters as fundamental frequency, word duration, vowel/nasal consonant spectra, voice onset time and glottal source spectrum slope. 
The parameters were estimated at manually identified locations from speech events within utterances. 
A later study~\cite{Zetterholm2001}, on a professional impersonator and one of his voice impersonations, showed that the impersonator not only focused on the voice of the target, but also matched the speech style and intonation pattern as well as the accent and pronunciation peculiar to the target. 

A voice impersonator may be identified by finding the features a typical impersonator chooses to exploit and what he ignores in the targeted voice, as had been tried in~\cite{Zetterholm2007} whereby two professional and one amateur impersonators were asked to mimic the same target in order to observe whether they have chosen the same features to change with the same degree of success.
The work described in~\cite{Kitamura2008} used professional impersonators\footnote{One interesting aspect of mimicry detection, in addition to employing professionals, could be to employ twins~\cite{rosenberg1976}, at least of reference} to mimic a person’s voice to identify the acoustic characteristics that each impersonator attempts to change to match the target. 
A comparison of the impersonated voices and the actual voice of the impersonator affirmed the importance of the pitch frequencies and vocal/glottal acoustics of the target speaker and impersonator. 
A similar work in~\cite{Amin2014}, involving three voice impersonators with nine distinct voice identities, recorded synchronous speech and Electro Glotto Graphic (EGG) signals. 
An analysis based on the EGG and the vocal traces - including speech rate, vowel formant frequencies, and timing characteristics of the vocal folds - led to the conclusion that each impersonator modulated every parameter during imitation. 
In addition, vowel pronunciations were observed to have a high dependency on the vowel category. 

More recently, the method in~\cite{Mary2012} uses a Support Vector Machine (SVM) to create speaker models based on the prosodic features (intonation, loudness, pitch dependent rhythm, intensity and mimic duration in addition to jitter, shimmer, energy change, and various duration measures) from the original speech of celebrities and professional mimicry artists; as well as the original speech of the latter. 
A related effort~\cite{Renjith2013} uses Bayesian interpretation in combination with SVM. 
A similar prosodic features-based method~\cite{Farrus2010}, analyzes the ability of impersonators to estimate the prosody of their target voices while using both intra-gender and cross-gender speeches.
%


\section{The Dataset}\label{sec:dset}
It seems that a standard dedicated speech impersonation database may not be publicly available, e.g., the study reported in ~\cite{campbell1997} used the YOHO database that was designed for ASV systems. 
The best one can get is to collect from online sources, like YouTube, audios of celebrities and their mimicked versions by various professionals. 
Alternatively, one may exploit the public datasets, like voxceleb~\cite{Nagrani2020} that contain the original voices of celebrities; one may still vie for human mimicked version voices of these celebrities on YouTube. A related dataset is~\cite{Mandalapu2021}. 

To collect the raw data, we went through a number of social media apps and sites and downloaded the audios which were then edited to conform to the proposed model by limiting it to a maximum duration of $20$ seconds in WAV format. 
One part of the dataset consists of all English audios (both real and mimicked). 
The second part of the dataset contains a mix of both English and Arabic audios. 
The audio files are named so that the first four characters are digits to represent the index and the fifth character is either 'r' or 'f' to label the voice as real or faked, respectively.

Our goal is to blindly identify whether a given voice is mimicked or otherwise. 
Hence, for our experiments, a set of independent real and faked audios was required to create the dataset; real and faked voices uttered independently of what is being said and who said it and independent of the language. We employed 933 distinct English spoken audio samples, divided into 746 for training (including $20\%$ for cross validation) and 187 for testing. To ensure language versatility, an additional $194$ Arabic medium samples were included, resulting in a total of $1127$ samples (901 for training and cross validation, and 226 for testing).
  
\section{The Proposed Method}\label{sec:method}
The use case of our method concerns the scenario of a complete blindness wherein no prior  or side information is available about the speaker. The idea is to accept/reject an input voice, right at the outset, without any recourse to the already available record. Our emphasis is on classifying the speech, as faked, or otherwise, under the assumption that no other recorded voice of the speaker is available; whether genuine or disguised.  The purported speaker is only represented once in the training data; that too, either as real or mimicked, but not both.  Potentially, our method may be very useful in improving efficiency of many audio processing methods, especially, when applied at the pre-processing stage.

The method we used is loosely based on the one described in~\cite{Vasconcelos2019} for recognizing the spoken digits (0--9) from the audio samples of six people. 
%
    \begin{figure*}[h!]
    \centering
   \includegraphics[width=5.0in]{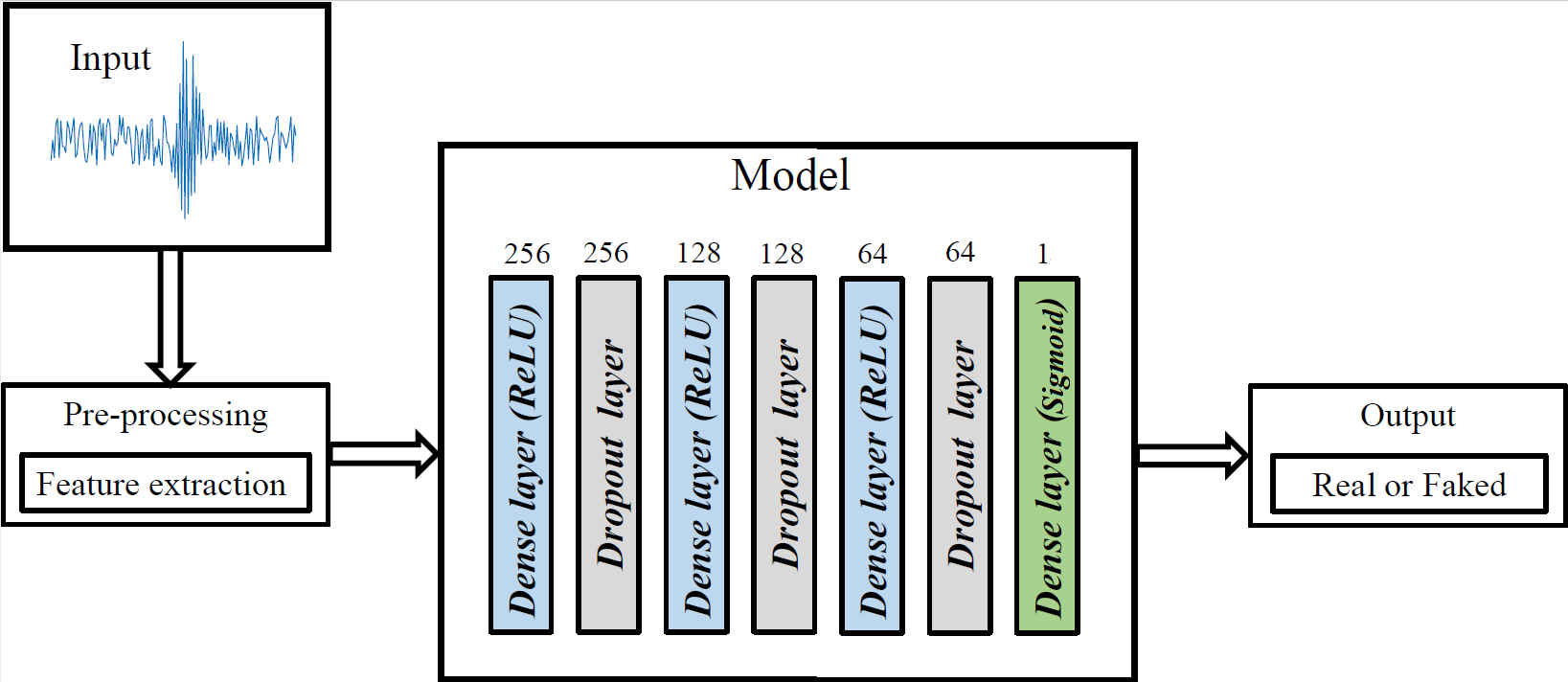}
   \caption{The proposed model.}
\label{fig:my_label2}
\end{figure*}
The proposed model is outlined in Fig.~\ref{fig:my_label2} that follows the steps given below.

\subsection{Input}
For the training set, it is essential that:
\begin{enumerate}
    \item Exactly  one sample pertains to a purported speaker.
    \item Either the real or the faked voice of a given speaker is part of the dataset; in fact, they should be mutually exclusive.
    \item  The spoken words are not required to be identical.
\end{enumerate}
Keeping the above in view, We extracted our dataset from the raw dataset to contain 
$933$ English-only audio samples and $1127$ samples in both  English and Arabic.

The input is constituted by the WAV files corresponding to the above that are stored in a suitable data structure. 
The model works in labeling our data based on the last letter ('f' or 'r') in the name of the file before storing it in a separate array of labels. 
The array is mapped to a separate perspective array of features obtained after the subsequent two steps.
    
\subsection{Feature extraction}
 The model works on extracting the needed features using the Python \texttt{librosa}~\cite{brian_mcfee_2022_6759664} package. 
 The main features we relied on were $26$ in total: 
     
\begin{itemize}
\item RMSE (E for Energy),
\item Zero crossing rate, 
\item Spectral centroid, 
\item Spectral bandwidth, 
\item Roll off, 
\item Chroma, 
\item $20$ MFCCs.
\end{itemize}

These features are already described in Section \ref{sec:speech} with some details. All these $26$ features would be normalized and concatenated, before being fed to the neural network, as input, during the training phase. 

\subsection{Pre-processing the dataset}
Using \texttt{sklearn.model\_selection}, the feature set is first partitioned to training and testing sets. 
During the training phase, the training set is dynamically partitioned to training and validation parts, as is the case with cross validation. 
By employing the \texttt{sklearn.preprocessing.StandardScaler} class, the data is normalized in order to better structure it for visualization and analyses. 
The data is standardized, which means that they will have a mean of 0 and a standard deviation of 1. 

\subsection{The Neural network}
The next step is the deep neural network outlined in Fig.~\ref{fig:my_label2} that follows a \textit{sequential model} that has three hidden layers.  
The reason to choose the sequential model is its simplicity and the ability to add up more layers easily. 

\begin{figure*}[h!]
             \centering
             \includegraphics[width=5in]{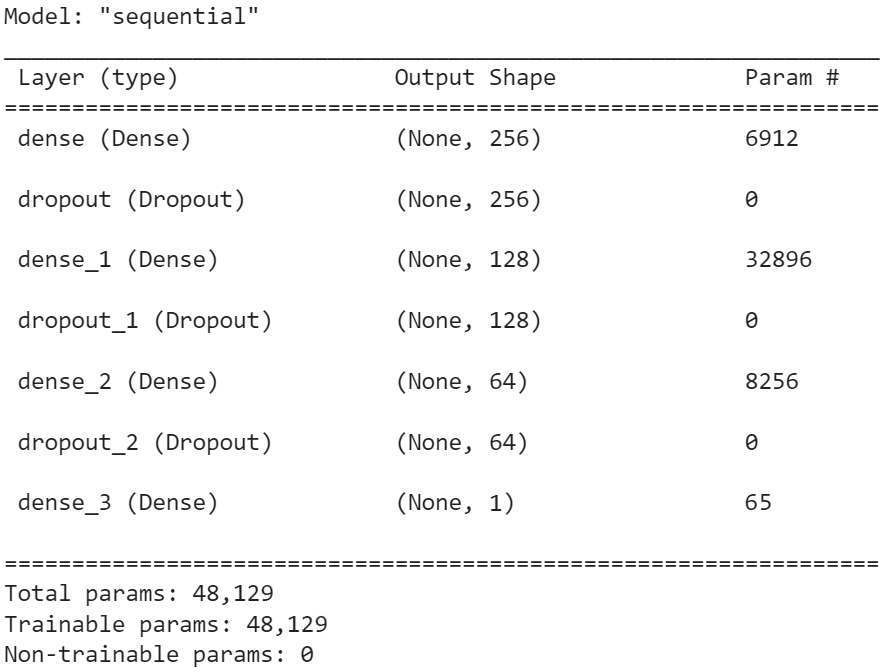}
             \caption{Model Summary}
             \label{fig:layers}
         \end{figure*}

The model has alternating dense and dropout layers.  
The dropout layer is used for toning down too many feature associations during training in order to avoid over-fitting; a phenomenon called regularization. The relevant hyperparameter $p$, called 'dropout rate', is set to $0.5$. 
For hidden layers, in general, the Rectified Linear Unit function (ReLU) is the activation function of choice. 
Being a binary classification, for the output layer, we are relying on the sigmoid activation function. We chose the sigmoid activation function for binary classification due to its ability to provide clear class probabilities between 0 and 1. While softmax is typically used for multi-class classification, sigmoid is well-suited for binary outcomes.  
Note that:
\begin{equation}
    ReLU(x) = max(0, x),
\end{equation}
\begin{equation}
    \text{Sigmoid}(x) = \frac{1}{1\ + e^{-x}}
\end{equation}
Fig.~\ref{fig:layers} gives a snapshot of the layers involved in an example execution in the form of model summary. 

After preparing the model, we pass on the data for training the model via the usual \texttt{fit()} function. 
We  use the adaptive moment estimation (\textit{adam}) as optimizer~\cite{Geron2019} for its efficiency, manageable memory requirements and its amenability for larger data/parameters. 
As we have only two possible labels (real or faked), for better accuracy, the loss function is based on the sparse categorical cross entropy. 
The number of epochs were set to 140 which is the number of times the model will train before it completes the training process. 
The batch-size was set to 128 and corresponds to the number of samples processed before the model is updated. 
Finally, the testing phase involved the usual prediction function (\texttt{predict()}) with subsequent comparison of the predicted labels with the actual labels from the dataset. The learning rate was optimized to $0.0003$.

\section{Experimental Results}\label{sec:results}
It must be noted that although the size of our dataset is almost double  the sample size we are reporting (every voice had a faked and real versions), for the sake of blindness either the real or fake version was included in the dataset. In other words, our sample has either a real or mimicked version of an individual's voice, but not both.

\begin{enumerate}
    \item The model was first trained and tested with all-English audios already described in the previous section. 
It consisted of $933$ samples which were partitioned into 
$746$ training ($20\%$ cross validation)  and $187$ test cases. 
The resultant confusion matrix corresponding to the training of the model on 
$933$ samples was:
   \begin{equation}
   Train\ (all\ English)\ \  
\begin{pmatrix}
TP & FP \\
FN & TN 
\end{pmatrix}
=
\begin{pmatrix}
372 & 13 \\
4 & 357 
\end{pmatrix},
\end{equation}
where:
     %
\begin{itemize}
\item  $TP$ = True positives,
\item  $TN$ = True Negatives,
\item  $FP$ = False Positives,
\item $FN$ = False Negatives.
\end{itemize}
%
%
Over the same dataset, the resultant confusion matrix par rapport the testing, based on $187$ samples, was observed to be:

\begin{equation}
Test\ (all\ English) =
    \begin{pmatrix}
    85 & 9 \\
    2 & 91 
    \end{pmatrix}.
\end{equation}

 
Based on the above confusion matrices, for all English dataset, the resultant training accuracy was found to be $0.977 (97.7\%)$ as against the testing accuracy of $94.1\%$. 
These results are interesting in the face of the fact that the method is blind and no additional information is made available to our method.
    \item To check whether the spoken language has any bearing upon the results, we chose a test case of $194$ Arabic language audios from the mixed set of data taken from the raw dataset. 
    When tested on the model trained on the English-only datset, we got a classification accuracy of $59.9\%$. 
    This implied that the spoken language also seems to be a deciding factor. We therefore ought to revisit the training part using a mix of English and Arabic audios. 
    In fact we combined the three sets to get our mixed set to retrain our network on. It must be noted that our use case has one and only one sample per purported speaker; 'real' or 'fake' as our use of the terms \textit{blind} and \textit{zero prior}  knowledge dictates. That is to say, if there's a 'fake' speech sample of a given purported speaker, then there's no 'real' speech sample, and vice versa. Neither has the proposed system ever encountered the purported speaker, whether fake or real. With that in mind, we must have no spoken language barrier in training our network. That is why we combined the English and Arabic speech samples in the mixed data set to nullify language dependence in our use case.
    \item  The mixed set of $1127$ audios was partitioned into $901$ training samples, including $20\%$ for cross validation, and the rest $226$ constituted the test set. 
The resultant confusion matrix was:
\begin{equation}
Train\ (Mixed) = 
    \begin{pmatrix}
450 & 13 \\
9 & 429 
\end{pmatrix}.
\end{equation}
%
The computed training accuracy was thus found to be 
$97.6\%$ in classifying the real and faked audios. 
The test accuracy for the mixed part was however 
$94.2\%$ as can be deduced from the following confusion matrix:
 \begin{equation}
Test\ (Mixed) = 
    \begin{pmatrix}
108 & 10 \\
3 & 105 
\end{pmatrix}.
\end{equation}

  

\end{enumerate}
 \begin{figure}[h!]
    \centering
    \subfloat[the English dataset]{\includegraphics[width=9cm]{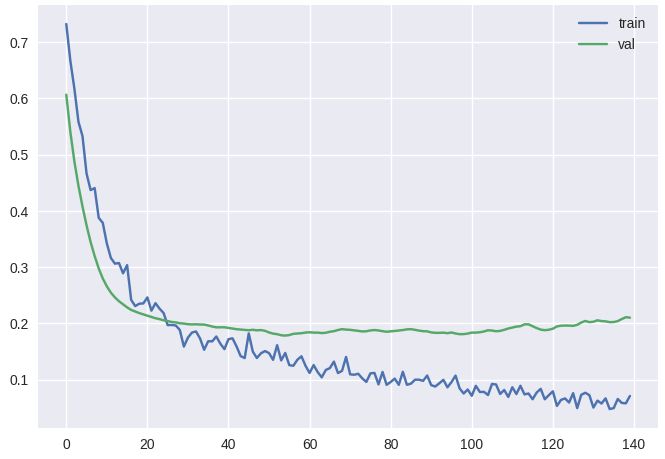}}\quad
    \subfloat[the mixed dataset]{\includegraphics[width=9cm]{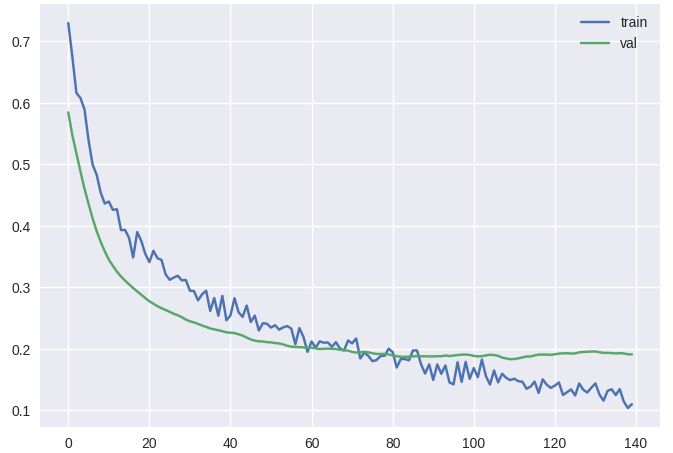}}\\
    \caption{Training and testing losses vs the number of epochs.}
    \label{fig:losses}
\end{figure}
 %
Fig.~\ref{fig:losses} indicates that our model has a good learning curve, albeit noisy at later part.
Note that we optimized based on the mixed data; hence the better curve (Fig.~\ref{fig:losses}.b).
 Although we fixed the number of epochs to $140$, most of the convergence is realized well before $60$ epochs; only some refining need further epochs. 
 The curves of loss decrease to a point of stability, although one can observe a small gap of validation loss with the training loss.
\begin{table*}[b!]
\scalebox{0.62}{
\tabulinesep=.3mm
   \begin{tabu}{|l|l|l|l|l|l|l|}
\cline{3-6}
\multicolumn{2}{l|}{}  & \multicolumn{2}{l|}{\textbf{All English Dataset (933)}} & \multicolumn{2}{l|}{\textbf{The Mixed Dataset (1127)}}\\ \hline
\cline{3-6}
\textbf{Metric}   &        \textbf{Formula}                      & {Train. (746)}  & {Test (187)}& {Train. (901)}  & {Test (226)} \\ \hline \hline
{\begin{tabular}{@{}c@{}}Sensitivity/Recall/\\True positive rate (TPR)\end{tabular}}   &  $\frac{TP}{TP+FN}$  & 0.989 & 0.977  & 0.980 & 0.973   \\ \hline
{\begin{tabular}{@{}c@{}}Specificity/\\True negative rate (TNR)\end{tabular}}  & $\frac{TN}{TN+FP}$  & 0.965  & 0.910  & 0.971  & 0.913 \\ \hline
{\begin{tabular}{@{}c@{}}Fall out/\\False positive rate (FPR)\end{tabular}}    & $\frac{FP}{FP+TN}$ & 0.035  & 0.09   & 0.029 &    0.087\\ \hline
{\begin{tabular}{@{}c@{}}Miss rate/\\False negative rate (FNR)\end{tabular}}   & $ \frac{FN}{FN+TP}$  & 0.011  & 0.023  & 0.02 &  0.027\\ \hline
{\begin{tabular}{@{}c@{}}Precision/\\Positive predictive value (PPV)\end{tabular}} & $\frac{TP}{TP+FP}$ & 0.966  & 0.904 & 0.972 &   0.915   \\ \hline
{Accuracy}   &  $\frac{TP+TN}{FN+TP+FP+TN}$  & 0.977 & 0.941 & 0.976 &    0.942 \\ \hline
{balanced accuracy (BA)}       & $\frac{Sensitivity\ +\ Specificity}{2}$ & 0.977 & 0.944 & 0.976 &  0.943   \\ \hline
{$F_1$ Score}       & $2\times \frac{Precision \times Recall}{Precision\ +\ Recall}$ & 0.977 & 0.939 & 0.925  & 0.976 \\ \hline
\end{tabu}}
\caption{\centering Metrics calculations for both the English and Mixed parts of the dataset.}
\label{tab:results}
\end{table*}

\begin{figure}[h!]
    \centering
    \subfloat[The English dataset]{\includegraphics[width=9cm]{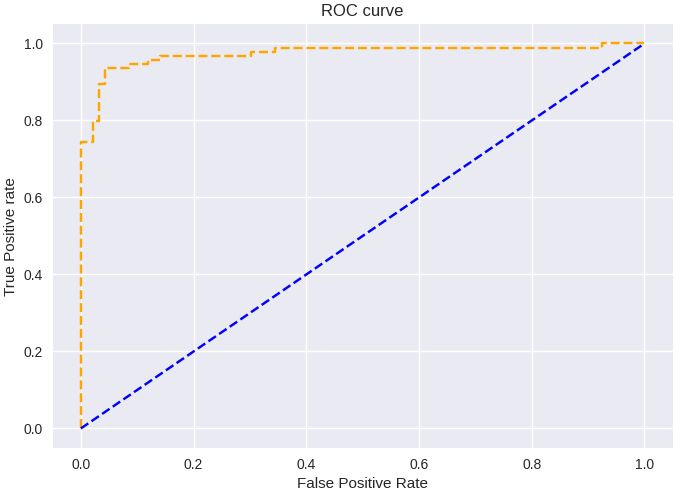}}\quad
    \subfloat[The mixed dataset]{\includegraphics[width=9cm]{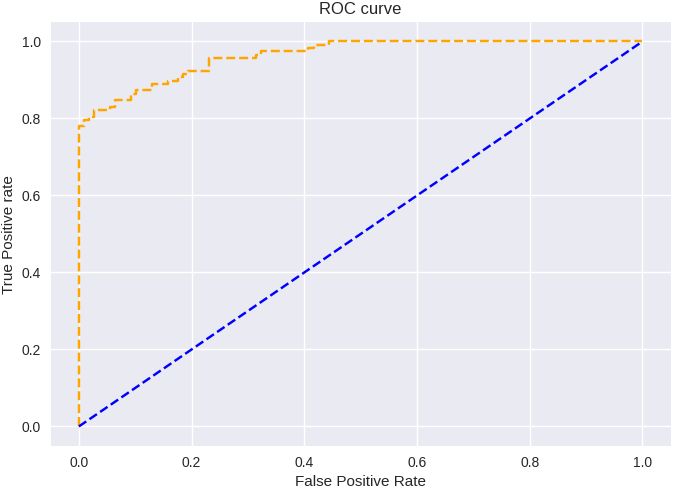}}\\
    \caption{Receiver Operating Characteristic (ROC) curves.}
    \label{fig:ROC}
\end{figure}

Table~\ref{tab:results} sums up all the results par rapport both the all English and mixed parts of the dataset. 
For a better idea about the obtained results, the Receiver Operating Characteristic (ROC) curves for both the parts are illustrated in Fig.~\ref{fig:ROC}. 
The model has enviable efficiency as demonstrated by the high ROC AUC scores of $0.974$ and $0.963$ with respect to All-English and Mixed samples, respectively. 
A very important ROC based metric is the equal error rate (EER) which is the location on a ROC 
curve where the false acceptance rate and false rejection rate are equal. In general, a lower EER  indicates highly accurate classification. The EER\footnote{\url{https://stackoverflow.com/questions/28339746/equal-error-rate-in-python}. Accessed: Jun. 07, 2023} in our case was $0.053$ for English-only and $0.057$ for the mixed part. 

\section{Benchmarking Results}\label{sec:ref}
Even after a thorough search, we were not able to find a reference method that could benchmark the research problem we are after. 
There are methods, like~\cite{Rodriguez-Ortega2020}, but they are non-blind, i.e., to say these methods require a recording of the original voice.  
One may argue in favor of spoofing detection literature, especially the one-class methods, like OC-Softmax~\cite{Zhang2021} and its variants~\cite{Li2023,Ding2023,Lin2024}, but the main method concerns machine generated speech. 
Hence, it was decided to come up with reference data by inspection through human subjects/volunteers. 
We gathered a group of 10 native Arabic speakers and 20 native English speakers. 
Each volunteer would have to listen the audios from our dataset and tabulate it as real or faked as per his/her observation. 
Due to the scarcity of native English speakers, we had to contact the volunteers through social media and carry out the process live online.

By conducting the tests on two groups of people, the English native speakers were able to blindly identify 85\% of the English audios given as real or fake correctly. 
In contrast, the proposed model got $94\%$ accuracy in classifying the real and faked audios using the all-English dataset. 
With the native Arab speakers, 89\% of the Arabic audios were identified correctly, however.

After conducting the human subject test and the results of the model test we found that there were some audios on which both tests agreed being faked audios, where in fact those audios were real, see Table\ref{peculiars}. 
The probable cause, of the failure of the test participants in identifying those audios, may be the background noise that may have made them think that those audios were real. 
\begin{table*}[!htbp]
\footnotesize
\centering
\begin{tabular}{|l|l|l|l|l|}
\hline
   \textbf{Language} & \textbf{Audio Id.} & \textbf{Human observers} & \textbf{Proposed Method} & \textbf{Ground Truth }\\ \hline
{Arabic}   & 0920f                        & real                                                                                 & real                                                                         & faked                                                                          \\ \cline{2-5} 
                      & 0931f                        & real                                                                                 & real                                                                         & faked                                                                          \\ \cline{2-5} 
                         & 0932f                        & real                                                                                 & real                                                                         & faked                                                                          \\ \hline
{English} & 0001f                        & real                                                                                 & real                                                                         & faked                                                                          \\ \cline{2-5} 
                         & 0003f                        & real                                                                                 & real                                                                         & faked                                                                          \\ \cline{2-5} 
                         & 0005f                        & real                                                                                 & real                                                                         & faked                                                                          \\ \cline{2-5} 
                         & 0006r                        & faked                                                                                 & faked                                                                         & real                                                                          \\ \cline{2-5} 
                         & 0007f                        & real                                                                                 & real                                                                         & faked                                                                          \\ \cline{2-5} 
                         & 0008f                        & real                                                                                 & real                                                                         & faked                                                                          \\ \cline{2-5} 
                         & 0009f                        & real                                                                                 & real                                                                         & faked                                                                          \\ \cline{2-5} 
                         & 0010f                        & real                                                                                 & real                                                                         & faked                                                                          \\ \cline{2-5} 
                         & 0011f                        & real                                                                                 & real                                                                         & faked                                                                          \\ \hline\\
\end{tabular}
\caption{\centering Matching wrong results by both the model and human observers.}
\label{peculiars}
\end{table*}
 



\section{Conclusion}\label{sec:concl}
The results reveal that, assuming zero prior  knowledge about the speaker and his speech, our system can classify a given speech as faked or otherwise on the fly. 
We had at our disposal only English and Arabic audios, but still we were able to deduce that the nature of spoken language may be important as we got less than $60\%$ accuracy in classifying via a network trained by English-only samples. 
That is why, subsequently, we trained on samples from both English and Arabic. This dramatically improved the results of the classification, notwithstanding Arabic constituted approximate $18-20\%$ of the dataset.  
The performance of our model is proven by its accuracy which was at least $94\%$ (to be exact, the test accuracy was $94.1\%$ for English dataset and $94.2\%$ for the mixed dataset as per Table \ref{tab:results}). 
A Comparison with results by inspection from human subjects proves that our model can identify real and faked audios with a better accuracy.

As a future improvement, we first aim to collect more data to improve the Arabic dataset and make it available for researchers. 
Secondly, there is a need of diversity in the form of the inclusion of audios in other languages too. 
This may improve the classification capability of the model. Last but not least, deploying the model in mobile based software may help against impersonation offenses.
Fundamentally,  for an audio clip to be classified with high accuracy as being fake, without any references, original, mimicked, or impersonated target even exist, a much deeper analysis will be needed to provide a plausible feature-set upon which such a decision is being made by the model.  
%


\end{document}